# Thermal Instability of Advection-Dominated Disks against Local Perturbations

Shoji Kato

*Department of Astronomy, Kyoto University, Sakyo-ku, Kyoto, 606-01*

and

Marek A. Abramowicz and Xingming Chen

*Department of Astronomy and Astrophysics, Göteborg University*

*and Chalmers University of Technology, 412 96 Göteborg, Sweden*

## Abstract

Thermal instability is examined for advection-dominated one-temperature accretion disks. We consider axisymmetric perturbations with short wavelength in the radial direction. The viscosity is assumed to be sufficiently small for the vertical hydrostatic balance to hold in perturbed states. The type of viscosity is given either by the $\alpha$-viscosity or by a diffusion-type stress tensor. Optically thick disks are found to be in general more unstable than optically thin ones. When the thermal diffusion is present, the optically thin disks become stable, but the optically thick disks are still unstable. The instability of the advection-dominated disks is different from that of the geometrically thin disks without advection. In the case of no advection, the thermal mode behaves under no appreciable surface density change. In the case of advection-dominated disks, however, the thermal mode occurs with no appreciable pressure change (compared with the density change), when local perturbations are considered. The variations of angular momentum and of surface density associated with the perturbations lead to a thermal instability. The astrophysical implications of this instability are briefly discussed.

## 1. Introduction

Thermal instability of geometrically thin accretion disks has been extensively studied with fruitful applications to many time varying phenomena in dwarf novae, X-ray stars, etc. The mechanism of this thermal instability is well understood, and its essential part can be explained by considering perturbations which occur without changing the surface density of the disks (e.g., Pringle 1976). The fact that this is a good approximation is closely related to the fact that the disk is geometrically thin. In geometrically thin disks, the radial force balance is realized by the balance between the centrifugal and gravitational forces. Because both forces are much stronger than the pressure force, a temperature change associated with the thermal mode occurs with no motion in the horizontal plane. The gas expands or shrinks only in the vertical direction so that a hydrostatic balance in the vertical direction is realized without changing the surface density. This approximation is valid even when perturbations are local in the radial direction.

In advection-dominated disks, however, the situation is different. Optically thick disks of this type have been introduced by Abramowicz, Lasota, & Xu (1986) and studied in details by Abramowicz et al. (1988), Kato, Honma, & Matsumoto (1988), and Chen & Taam (1993). Optically thin advection-dominated disks have been most recently investigated independently by Abramowicz et al. (1995) and Narayan & Yi (1994, 1995a,b) and by Chen et al. (1995). [It is noted, however, that optically thin advection-dominated transonic disks are already introduced by Matsumoto, Kato, & Fukue (1985), although full details are not published.] Advection-dominated disks are not vertically thin since the local disk half-thickness $H(r)$ is close to the radial coordinate $r$ itself (Abramowicz et al. 1988, 1995; Narayan & Yi 1995a). This means that the pressure force in the radial direction is non-negligible compared with the centrifugal and gravitational forces. Therefore a pressure change associated with the thermal perturbation brings about a motion in the radial direction. This results a variation of surface density which is larger than the pressure variation when a local perturbation is considered, as will be discussed later. Thus, the approximation of no change of surface density is no longer valid and the criterion of thermal instability becomes quite different from that in the case of geometrically thin disks, especially when the local perturbation is considered. This difference has not been fully recognized in the previous studies of thermal instability of advection-dominated disks (Wallinder 1990; Honma et al. 1991; Chen & Taam 1993). A purpose of this paper is to clear this difference and to derive the instability criterion against local perturbations. Local perturbations mean here that the radial wavelength of perturbations is shorter than $r$, where $r \geq H$.

In addition to the above formal interest in thermal instability of advection-dominated disks, this kind of instability may have important astrophysical applications. It was recently recognized that advection-dominated disks are possible models of some AGN (Abramowicz et al 1995), of X-ray transients (Narayan & Yi 1995), and of the Sgr A$^*$ (Narayan, Yi, & Mahadavan 1995). Some of the time variations observed in these objects will be related to the thermal instability of advection-dominated disks.

## 2. Vertically Integrated Equations

We consider axisymmetric perturbations on axisymmetric disks. To this examination we introduce a cylindrical system of coordinates $(r, \varphi, z)$ which is centered on the central object and its $z$-axis is chosen as the rotation axis of the disk.

## 2.1. Validity of Vertically Integrated Equations

We focus here the advection-dominated accretion disks. Considering that $H/r$ is of order of unity and also the presence of the fast radial flow motion, we must be careful about the validity of using the vertically integrated equations which are widely used in the analyses of thermal instability of geometrically thin disks.

To examine the validity condition we compare here the time scale associated with perturbations with the dynamical time scale in which the vertical hydrostatic balance is realized. The growth rate of the thermal instability in geometrically thin $\alpha$-disks is of the order of $\alpha\Omega$, where $\Omega$ is the angular velocity of disk rotation. However, as we show later, in advection-dominated disks, for perturbations with short radial wavelengthes, the growth rate becomes $\alpha\Omega A$. The factor $A$ depends on the model of the dissipation processes. In the case which will be discussed in section 4, we have $A = (kr)^{1/2}(H/r)$, where $A = 1$ and $A = (kH)^2$ in the cases discussed in sections 5.1 and 5.2, respectively. Here $k$ is the radial wavelength of perturbations. In addition to this, we must consider the time variation due to the propagation of the perturbation. The frequency associated with this is $kU$, where $U(r)$ is the radial velocity of advection flow,

$$U \sim \alpha \Omega r \left(\frac{c_{\rm s}}{\Omega r}\right)^2, \tag{1}$$

and $c_{\rm s}$ is the local sound speed. These frequencies must be lower than the dynamical frequency, $c_{\rm s}/H (=\Omega)$, in order that one may use the vertically integrated equations since they are derived by assuming hydrostatic balance. The condition is

$$\alpha \Omega A < \Omega \quad \text{and} \quad kU < \Omega, \tag{2}$$

and can be rewritten as

$$\alpha < A^{-1} \quad \text{and} \quad \alpha < (kH)^{-1}\frac{r}{H}. \tag{3}$$

Inequality (3) shows that local perturbations, $kr \gg 1$, can be studied using the vertically integrated equations as long as $\alpha$ is sufficiently small. In this paper we restrict our discussion to this case.

## 2.2. Basic Equations

We assume the disk is axisymmetric and non-self-gravitating. The basic equations which we adopt here are the time-dependent version of those given by Matsumoto et al. (1984). That is, the basic hydrodynamic equations are integrated in the vertical direction under the assumption that a hydrostatic balance always holds in the vertical direction. To avoid ambiguities regarding the vertical integration, the integration is carried out under the assumption that the pressure, $p$, is related to the density, $\rho$, through the polytropic relation, i.e., $p \propto \rho^{1+1/N}$, with $N = 3$ for simplicity. The effects of general relativity are introduced by the pseudo-Newtonian potential, $\psi = -GM/(R - r_g)$, of Paczyński & Wiita (1980). Here $R$ is the distance from the central object of mass $M$ and $r_g = 2GM/c^2$ is its Schwarzschild radius. Under these assumptions, the equation of continuity is

$$\frac{\partial \Sigma}{\partial t} + \frac{1}{r}\frac{\partial}{\partial r}(r\Sigma U) = 0, \qquad (4)$$

where $\Sigma$ is the surface density obtained by the vertical integration of $\rho$, and $U(<0)$ is the radial component of the velocity. Hereafter, $U$ is taken to be negative, although $U$ used before was the absolute value of the radial flow. The $r$- and $\varphi$- components of equation of motions are, respectively,

$$\frac{DU}{Dt} - \frac{V^2}{r} + \Omega_{\rm K}^2 r + \frac{1}{\Sigma}\frac{\partial W}{\partial r} + \frac{W}{\Sigma}\frac{d\ln\Omega_{\rm K}}{dr} = 0, \qquad (5)$$

and

$$\frac{DV}{Dt} + \frac{UV}{r} + \frac{1}{r^2 \Sigma}\frac{\partial}{\partial r}(r^2 W_{r\varphi}) = 0, \qquad (6)$$

where $W$ is the vertically integrated $p$, $V$ is the $\varphi$-component of the velocity, and $\Omega_{\rm K}(r)$ is the Keplerian angular velocity on the equatorial plane, i.e.,

$$\Omega_{\rm K} = \left(\frac{1}{r}\frac{\partial \psi}{\partial r}\right)^{1/2}\bigg|_{z=0}. \qquad (7)$$

The operator $D/Dt$ is defined as

$$\frac{D}{Dt} = \frac{\partial}{\partial t} + U\frac{\partial}{\partial r}. \qquad (8)$$

In the case of $N = 3$, the vertical integration gives (Hoshi 1977)

$$\Sigma = 2I_3 \rho_c H, \qquad W = 2I_4 p_c H, \qquad (9)$$

where $H$ is the half disk-thickness, $\rho_c$ and $p_c$ are the density and the pressure on the equatorial plane respectively, and $I_3 = 16/35$ and $I_4 = 128/315$ are numerical constants.

The quantity $W_{r\varphi}$ in equation (6) is the vertical integral of the $r\varphi$-component of the viscous stress tensor. We adopt here the standard $\alpha$-viscosity (Shakura & Sunyaev 1973)

$$W_{r\varphi} = \alpha W. \tag{10}$$

The hydrostatic balance in the $z$-direction gives

$$(\Omega_{\rm K} H)^2 = 2(N+1)\frac{p_c}{\rho_c}. \tag{11}$$

In the case of $N = 3$, the vertical integration of energy equation can be written in a simple form (Matsumoto et al. 1984):

$$\frac{I_4}{I_3}\Sigma T_c \frac{DS_c}{Dt} = Q^+ - Q^-. \tag{12}$$

Here, $T_c$ and $S_c$ are the temperature and the specific entropy of the gas on the equatorial plane respectively, $Q^+$ is the heating rate per unit surface given by

$$Q^+ = -rW_{r\varphi}\frac{\partial}{\partial r}\left(\frac{V}{r}\right), \tag{13}$$

and $Q^-$ is the radiative flux escaping from the disk surface. The detailed expression for $Q^-$ will not effect the advection-dominated disks and thus is omitted here. Finally, the equation of state for optically thin disks is

$$p_c = \frac{k}{\mu m_{\rm H}}\rho_c T_c. \tag{14}$$

For optically thick disks the radiation pressure, $(1/3)aT_c^4$, is added on the right-hand side of equation (14).

In section 5, some modifications of these equations are discussed: Instead of the standard $\alpha$-viscosity, a diffusion-type stress tensor is adopted. Furthermore, i) a viscous force is added to the right-hand side of equation (5) and ii) a thermal diffusion term is taken into account on the right-hand side of equation (12).

## 3. Equations of Perturbations under Geostrophic Approximation

A small-amplitude, short-wavelength perturbation is superimposed onto an unperturbed disk. At the radius where local perturbations are imposed, the unperturbed disk is assumed to be rotating cylindrically, $\Omega = \Omega(r)$, and the radial velocity of the flow $U_0$ depends only on the radius, $U_0 = U_0(r)$. Eulerian perturbations superimposed over the unperturbed quantities $\Sigma_0$, $U_0$, $V_0(=\Omega r)$, and $W_0$ are denoted by $\Sigma_1$, $U_1$, $V_1$ and $W_1$, respectively. Hereafter we adopt dimensionless variables:

$$\sigma = \frac{\Sigma_1}{\Sigma_0}, \quad w = \frac{W_1}{W_0}, \quad u = \frac{U_1}{U_0} \quad \text{and} \quad v = \frac{V_1}{\Omega r}. \tag{15}$$

They are taken to be proportional to $\exp(n\Omega t - ikr)$, where $n$ is the dimensionless growth rate of perturbations normalized by $\Omega$. Hereafter, the subscripts 0 to $\Sigma$, $W$, and $U$ are omitted for simplicity.

From equation (4) we have

$$n_*\sigma - ikr\frac{U}{\Omega r}u = 0, \tag{16}$$

where

$$n_* = n - ikr\frac{U}{\Omega r}. \tag{17}$$

When considering the $r$-component of equation of motion, we introduce an approximation to filter out the acoustic waves. For low frequency perturbations in a rapidly rotating system, the pressure force is roughly balanced by the Coriolis force. That is, the term of $DU/Dt$ in equation (5) can be neglected. This approximate procedure is well known as the geostrophic approximation in meteorology and oceanography (e.g., Pedlosky 1979). (We can show later, using the final results, that the terms neglected are higher order small quantities.) Under this approximation, we have from equation (5)

$$-2v - ikr\left(\frac{c_s}{\Omega r}\right)^2 w - \frac{d\ln(W\Omega_K)}{d\ln r}\left(\frac{c_s}{\Omega r}\right)^2 \sigma = 0, \tag{18}$$

where $c_s$ is the speed of sound defined by $c_s^2 = W/\Sigma$. The $\varphi$-component of equation of motion (6) leads to

$$n_*v + \frac{\kappa^2}{2\Omega^2}\frac{U}{\Omega r}(u+\sigma) - i\alpha kr\left(\frac{c_s}{\Omega r}\right)^2 w = 0, \tag{19}$$

where $\kappa$ is the epicyclic frequency defined by $\kappa^2 = 2\Omega(2\Omega + rd\Omega/dr)$. Finally, energy equation (12) yields, after lengthly calculations,

$$n_*\Omega(C_1 w - C_2\sigma) = G_v v + G_w w + G_\sigma \sigma + G_u u, \tag{20}$$

where

$$G_v = i\alpha\Omega kr \tag{21a}$$

$$G_w = \left(\frac{\partial Q^+}{\partial W}\right)_{\Sigma,V} - \left(\frac{\partial Q^-}{\partial W}\right)_{\Sigma} - C_3 \frac{Q^-_{\text{ad}}}{W}, \qquad (21b)$$

$$G_\sigma = -\frac{1}{c_s^2}\left(\frac{\partial Q^-}{\partial \Sigma}\right)_W - C_4 \frac{Q^-_{\text{ad}}}{W}, \qquad (21c)$$

$$G_u = -\frac{Q^-_{\text{ad}}}{W}. \qquad (21d)$$

Here, $Q^+$ and $Q^-$ are the viscous heating and radiative cooling rates in the unperturbed state, and $Q^-_{\text{ad}}$ is the advection cooling rate in the unperturbed state, i.e.,

$$Q^-_{\text{ad}} = \frac{I_4}{I_3}\Sigma T_c \frac{DS_c}{Dt} = Q^+ - Q^-. \qquad (22)$$

The first term on the right-hand side of equation (20) comes from variation of heating due to a change of rotational velocity $V$ [see equation(13)]. In the above equations $C$'s are numerical constants defined by

$$C_1 = \frac{\Gamma_1 + 1}{2(\Gamma_3 - 1)}, \qquad C_2 = \frac{3\Gamma_1 - 1}{2(\Gamma_3 - 1)}, \qquad (23a)$$

and

$$C_3 = \frac{1+\beta}{2(4-3\beta)}, \qquad C_4 = \frac{9(1-\beta)}{2(4-3\beta)}, \qquad (23b)$$

where $\Gamma_1$ and $\Gamma_3$ are adiabatic indices which include the effects of radiation pressure, and $\beta$ is the ratio of the gas pressure to the total pressure.

In the case of optically thick, advection-dominated disks, we have $\gamma = 4/3$ and $\beta = 0$, since $p_r \gg p_g$. Then the constants defined above become

$$C_1 = \frac{7}{2}, \quad C_2 = \frac{9}{2}, \quad C_3 = \frac{1}{8}, \quad C_4 = \frac{9}{8}. \qquad (24a)$$

On the other hand, in optically thin, advection-dominated disks, we have $\gamma = 5/3$ and $\beta = 1$, since the disks are supported by ion pressure. Hence we adopt

$$C_1 = 2, \quad C_2 = 3, \quad C_3 = 1, \quad C_4 = 0. \qquad (24b)$$

### 4. Thermal Instability agaist Local Perturbations: $\alpha$ viscosity

Before examining in detail the thermal instability in advection-dominated disks, we shall discuss briefly the reason for thermal instability in geometrically thin disks. Comparison between these two is helpful to understand the essential difference of thermal instability in geometrically thick disks and geometrically thin disks.

## 4.1. Local Instability of Geometrically Thin Disks

It is well known that the thermal instability in geometrically thin disks can be examined under the approximation that there is no surface density change during the growth of the instability. Under this approximation, the $v$ and $\sigma$ terms in the energy equation (20) can be neglected. Furthermore, the $G_u u$ term in that equation can also be neglected, since $G_u$ is negligibly small in disks with no advective energy transport. Thus, energy equation (20) can be approximated as

$$n\Omega C_1 w = \tilde{G}_w w, \tag{25}$$

where

$$\tilde{G}_w = \left(\frac{\partial Q^+}{\partial W}\right)_{\Sigma,V} - \left(\frac{\partial Q^-}{\partial W}\right)_{\Sigma}. \tag{26}$$

Here, term $Q^-_{\rm ad}$ has been neglected in $\tilde{G}_w$, since the advective cooling is negligible in geometrically thin disks. The condition of the thermal instability is then

$$\tilde{G}_w > 0. \tag{27}$$

This is the well-known criterion for thermal instability (e.g., Pringle 1976). Equation (25) shows that the order of $n$ is $\alpha$.

If the $\sigma$ terms are neglected in the $r$- and $\varphi$-components of momentum equation, we have from equations (18) and (19)

$$-2v - ikr\left(\frac{c_{\rm s}}{\Omega r}\right)^2 w = 0, \tag{28}$$

and

$$nv + \frac{\kappa^2}{2\Omega^2}\frac{U}{\Omega r}u - i\alpha kr\left(\frac{c_{\rm s}}{\Omega r}\right)^2 w = 0. \tag{29}$$

These equations show that $O(v) = (kH)(H/r)O(w)$ and $O(u) = (kr)O(w)$, where $O$ denotes the order of the magnitude of the subsequent argument. Furthermore, the continuity equation (16),

$$n\sigma - ikr\frac{U}{\Omega r}u = 0, \tag{30}$$

gives $O(\sigma) = (kH)(H/r)O(u)$. Thus, with $kH \ll 1$ and $H/r \ll 1$, we have

$$O(v) \ll O(\sigma) \ll O(w) \ll O(u). \tag{31}$$

In this way, the neglection of the $\sigma$-terms in the momentum equation is found to be consistent with the final results. The above results also confirm the validity of approximations used to derive equation (25).

After obtaining $n$ from equation (25), we can derive $v$, $u$, and $\sigma$, successively, from equations (28) – (30), as functions of $n$ and $w$. We shall discuss now how the above situations are changed when local perturbations ($kr \gg 1$) are considered in advection-dominated disks ($H \leq r$).

*4.2. Local Perturbations in Advection-Dominated Disks*

Advection-dominated disks are not geometrically thin, $H/r \leq 1$. This means that the pressure force is non-negligible compared with the gravitational and centrifugal forces. Hence, if a temperature increment occurs at a certain localized region, the disk at the region expands in the radial direction. For a fixed magnitude of radial expansion, the surface-density decrease due to the expansion is larger in the case when radial wavelength is shorter. On the other hand, the decrease of vertically integrated pressure in that region due to the radial expansion is smaller when the radial wavelength is shorter. This follows from the fact that the pressure at a particular radius is determined by the weight of the mass supported by the effective gravitational force (the difference between the centrifugal force and the gravitational one) outside this radius. (This can be understood by integrating the $r$-component of equation of motion from $r = \infty$ to the radius in question.) In other words, the pressure can not change much (compared with the density change) by local perturbation since it does not change the global structure of the force balance. In summary, we can expect a large density change, but a small pressure change, when local perturbations are considered. That is, $O(w) < O(\sigma)$. This is a difference from the case of geometrically thin disks.

The fact that in local perturbations a large density change occurs with a small pressure change is well known in stellar hydrodynamics in relation to flickers. A flicker is a phenomenon which occurs by thermal instability of a localized perturbation in shell burning region of stars, found first by Schwarzschild & Härm (1965) (e.g., Fujimoto, Sugimoto 1979). The thermal instability of geometrically thick disks agaist local perturbations is similar to the flickers in the sense that $O(w) < O(\sigma)$, although the heating sources leading to instabilities are different in each case.

Let us now consider the orders of variables, $v$, $w$, $u$, and $\sigma$. In the present case the order of $n_*$ is $\alpha(kr)^{1/2}(H/r)$, as will be confirmed later. If this is adopted, the equation of continuity, equation (16), shows that the amplitude of $u$ is smaller than that of $\sigma$ by a factor of $(kr)^{-1/2}(H/r)^{-1}$, since the order of $U$ is given by equation (1). That is, we have $O(u) = (kr)^{-1/2}(H/r)^{-1}O(\sigma)$. This means that the $u$ term in equation (19) can be neglected, compared with the term of $\sigma$. Comparison of equation (18) and equation (19) shows further that the first term on the left-hand side of equation (18), $-2v$, can be neglected. [The $w$ and $\sigma$ terms in equation (18) have the orders of the corresponding terms in equation (19) times $\alpha^{-1}$. On the other hand, the $v$ term in equation (18) has the order of the corresponding term in equation (19) times $\alpha^{-1}(kr)^{-1/2}(H/r)^{-1}$, which is smaller than $\alpha^{-1}$]. Thus, equations (18) and (19) can be approximated, respectively, as

$$-ikr\left(\frac{c_s}{\Omega r}\right)^2 w - \frac{d\ln(W\Omega_K)}{d\ln r}\left(\frac{c_s}{\Omega r}\right)^2 \sigma = 0, \tag{32}$$

and

$$n_* v + \frac{\kappa^2}{2\Omega^2}\frac{U}{\Omega r}\sigma - i\alpha kr\left(\frac{c_s}{\Omega r}\right)^2 w = 0. \tag{33}$$

These are the $r$- and $\varphi$- components of equation of motion.

We have mentioned that in local perturbations in advection-dominated disks the order of $w$ is smaller than that of $\sigma$. Equation (32) really shows that the ratio is about $(kr)^{-1}$, i.e., $O(w) = (kr)^{-1}O(\sigma)$. Equation (33) shows then that $O(v) = (kr)^{-1/2}(H/r)O(\sigma)$. In summary, we have

$$O(w) < O(v) \sim O(u) < O(\sigma). \tag{34}$$

The use of the above relations among $w$, $v$, $u$, and $\sigma$ leads us to an approximate form of equation (20):

$$n_*\Omega(-C_2\sigma) = G_v v. \tag{35}$$

This is the energy equation. Combination of equations (32), (33) and (35) gives the final dispersion relation,

$$n_*^2 = \frac{\alpha}{C_2}\frac{G_v}{\Omega}\left(\frac{c_s}{\Omega r}\right)^2\left[\frac{d\ln(W\Omega_K)}{d\ln r} + \frac{\kappa^2}{2\Omega^2}\frac{U}{\alpha\Omega r(c_s/\Omega r)^2}\right]$$

$$= i\alpha^2(kr)\frac{1}{C_2}\left(\frac{c_s}{\Omega r}\right)^2\left[\frac{d\ln(W\Omega_K)}{d\ln r} + \frac{\kappa^2}{2\Omega^2}\frac{U}{\alpha\Omega r(c_s/\Omega r)^2}\right]. \tag{36}$$

This shows the presence of an unstable mode whose growth rate is of the order of $\alpha\Omega(kr)^{1/2}$ $(H/r)$. This growth rate is consistent with the initial assumption concerning the order of $n_*$.

The right-hand side of equation (35) shows that the direct source of the thermal instability is angular momentum variation. The angular momentum variation occurs as a natural consequence of density variation. If angular momentum per unit surface is conserved (in the Lagrangian sense) during the growth of perturbations, we have $[d(\Sigma V)/dt]_1 = 0$. This equation relates directly $v$ and $\sigma$ when the change of $U$, i.e., $u$, is neglected. That is, we have $n_* v + (\kappa^2/2\Omega^2)(U/\Omega r)\sigma = 0$. This is an approximate version of equation (33). If this equation and equation (35) are combined to derive a dispersion relation, we obtain an equation which is the same as equation (36) except that the first term in the brackets of equation (36) is omitted.

*4.3. Comments on the Case with no Shear*

As shown above, the main source of the thermal instability in geometrically thick $\alpha$-disks is the angular momentum variation associated with perturbations. However, consideration of the case when the effect of angular momentum variation is neglected seems to be instructive for understanding the relation of the present instability with that of non-rotating stars (i.e., flickers). If the effects of angular momentum variation is neglected, the main term on the right-hand side of equation (20) is only $G_\sigma \sigma$ since $G_w w$ and $G_u u$ can be neglected simply because $O(w), O(u) < O(\sigma)$. Then, one can obtain easily the growth rate as
$$n_*\Omega = -\frac{G_\sigma}{C_2}, \tag{37}$$
and the condition of instability is thus $G_\sigma < 0$.

This instability criterion can be understood as following. Let us consider a surface density decrease at a localized region. If $G_\sigma < 0$ (which is true in most of the cases in advection-dominated disks, see section 5), this density decrease leads to an entropy increase at the place, since the right-hand side of equation (20) represents entropy change. Since an entropy change is proportional to $C_1 w - C_2 \sigma \sim -C_2 \sigma$, the entropy increase results in a decrease of the surface density. Therefore, the surface density decreases further more. This instability mechanism is the same, in principle, as that of flickers in stars. (It should also be noted that in the present case the term of $-2v$ should be retained on the left-hand side of equation (32), but this makes no essential difference in the argument presented above.)

## 5. Thermal Instability against Local Perturbations: Diffusion-Type Viscosity

*5.1. The Case with no Thermal Diffusion*

The arguments in the previous section have some drawbacks. First, the results show that the spatial variations of angular momentum due to the perturbation are the main source for thermal instability. This suggests that we must re-examine the same problem in the case where the stress tensor has a diffusion form, because the stress tensor described by the $\alpha$-model does not reflect properly the effects of variations of angular momentum. As the $r\varphi$-component of the stress tensor we adopt here
$$W_{r\varphi} = -\eta r \frac{d}{dr}\left(\frac{V}{r}\right), \tag{38}$$

where $\eta$ is the vertically integrated turbulent dynamical viscosity. Its magnitude is about $\alpha_* \Sigma c_\mathrm{s}^2/\Omega$ and $\alpha_*$ is the dimensionless viscosity parameter corresponding to the $\alpha$ used before. If equation (38) is used instead of equation (10), we must change equation (19) to

$$n_* v + \frac{\kappa^2}{2\Omega^2} \frac{U}{\Omega r}(u + \sigma)$$

$$+ i\alpha_* kr \left(\frac{c_\mathrm{s}}{\Omega r}\right)^2 \left[\frac{d\ln\Omega}{d\ln r}\left(\frac{\partial\ln\eta}{\partial\ln W}\right)_\Sigma w + \frac{d\ln\Omega}{d\ln r}\left(\frac{\partial\ln\eta}{\partial\ln\Sigma}\right)_W \sigma - ikrv\right] = 0. \qquad (39)$$

Second, in the conventional $\alpha$-model the viscous force in the radial direction is not taken into account in the $r$-component of momentum equation. However, the force might have non-negligible effects on behavior of thermal perturbations, since we are considering perturbations whose wavelength in the radial direction is short. The main term which should be added on the right-hand side of equation (5) is $\eta\partial^2 U/\partial r^2$. Hence, equation (18) is changed to

$$-2v - ikr\left(\frac{c_\mathrm{s}}{\Omega r}\right)^2 w - \frac{d\ln(W\Omega_\mathrm{K})}{d\ln r}\left(\frac{c_\mathrm{s}}{\Omega r}\right)^2 \sigma = -\alpha_*(kr)^2\left(\frac{c_\mathrm{s}}{\Omega r}\right)^2 u. \qquad (40)$$

In addition to the above modifications, equation (20) must also be changed to

$$n_* \Omega(C_1 w - C_2 \sigma) = \bar{G}_v v + \bar{G}_w w + \bar{G}_\sigma \sigma + \bar{G}_u u, \qquad (41)$$

where

$$\bar{G}_v = -2ikr\left(\frac{d\ln\Omega}{d\ln r}\right)^{-1}\frac{Q^+}{W} \qquad (42a)$$

$$\bar{G}_w = \frac{Q^+}{W}\left(\frac{\partial\ln\eta}{\partial\ln W}\right)_\Sigma - \frac{Q^-}{W}\left(\frac{\partial\ln Q^-}{\partial\ln W}\right)_\Sigma - C_3 \frac{Q^-_\mathrm{ad}}{W}, \qquad (42b)$$

$$\bar{G}_\sigma = \frac{Q^+}{W}\left(\frac{\partial\ln\eta}{\partial\ln\Sigma}\right)_W - \frac{Q^-}{W}\left(\frac{\partial\ln Q^-}{\partial\ln\Sigma}\right)_W - C_4 \frac{Q^-_\mathrm{ad}}{W}, \qquad (42c)$$

$$\bar{G}_u = -\frac{Q^-_\mathrm{ad}}{W}. \qquad (42d)$$

The last basic equation, equation (16), remains unchanged. The set of equations to be used here is equations (39) – (41), and (16).

The main terms in equation (39) are those in the brackets, since they are multiplied by the factor $kr$. This means that the angular momentum transport by viscosity is so efficient that $\partial(r^2 W_{1r\varphi})/r^2 \partial r \sim 0$ holds. Furthermore, among the terms in the brackets, the last two are the dominant terms. Thus $O(v) = (kr)^{-1}O(\sigma)$. The final results which will be given below show that in the present case the order of $n_*$ is $\alpha_*$, i.e., $O(n_*) = \alpha_*$. When this is taken into account, we have $O(u) = (kr)^{-1}(H/r)^{-2}O(\sigma)$ from equations (16) and $O(w) \leq (kr)^{-1}O(\sigma)$ from equation (40). In summary, we have

$$O(v) \sim O(w) \sim O(u) < O(\sigma). \tag{43}$$

Hence, the basic equations (16), (40), (39), and (41), describing the local thermal perturbations, can be approximated, respectively, as

$$n_*\sigma - ikr\frac{U}{\Omega r}u = 0, \tag{16}$$

$$-ikrw - \frac{d\ln(W\Omega_K)}{d\ln r}\sigma = -\alpha_*(kr)^2 u, \tag{44}$$

$$\frac{d\ln\Omega}{d\ln r}\left(\frac{\partial\ln\eta}{\partial\ln\Sigma}\right)_W \sigma - ikrv = 0, \tag{45}$$

and

$$n_*\Omega(-C_2\sigma) = \bar{G}_v v + \bar{G}_\sigma \sigma. \tag{46}$$

The elimination of $v$ from equations (45) and (46) gives

$$n_*\Omega(-C_2\sigma) = G\sigma \tag{47}$$

where

$$G = -i\bar{G}_v \frac{1}{kr}\frac{d\ln\Omega}{d\ln r}\left(\frac{\partial\ln\eta}{\partial\ln\Sigma}\right)_W + \bar{G}_\sigma$$

$$= -\frac{Q^+}{W}\left(\frac{\partial\ln\eta}{\partial\ln\Sigma}\right)_W - \frac{Q^-}{W}\left(\frac{\partial\ln Q^-}{\partial\ln\Sigma}\right)_W - C_4\frac{Q^-_{\text{ad}}}{W}, \tag{48}$$

and the condition of instability is

$$G < 0. \tag{49}$$

The growth rate is given by $-G/C_2$ and which is of the order of $\alpha_*\Omega$ in consistent with the assumption made before. Equations (16) and (44) are subsidiary relations which express $u$ and $w$ in terms of $\sigma$ and $n_*$.

Among the three terms on the right-hand side of equation (48), those with $Q^+$ and $Q^-_{\text{ad}}$ are the main terms, since $Q^+ \sim Q^-_{\text{ad}} \gg Q^-$ in advection-dominated disks. In the conventional model of turbulent viscosity, we take $\eta \propto \Sigma c_s^2/\Omega \propto W/\Omega$ and thus $(\partial\ln\eta/\partial\ln\Sigma)_W \sim 0$. This means that the term of $Q^+$ is small and $G \sim \bar{G}_\sigma$. Hence, the growth rate is practically equivalent to that given by equation (37). In optically thick, advection-dominated disks, $\bar{G}_\sigma$ is negative since $C_4 > 0$. In the case of optically thin, advection-dominated disks, however, the major term of $\bar{G}_\sigma$ vanishes in the lowest approximations since $C_4 \sim 0$. This means that in the case of optically thin disks, some of the terms neglected in deriving equation (48) should be restored to derive the growth rate. However, the main effect of the inclusion of these terms is just adding an imaginary part to $n_*$. The real part of $n_*$ is still given by equation (48), unless $G$ is too small. Hence, the instability criterion is still given by $G < 0$ (or $\bar{G}_\sigma < 0$), and the disks are unstable, since $(\partial \ln Q^-/\partial \ln \Sigma)_W > 0$ for the free-free cooling.

### 5.2. Effects of Thermal Diffusion

In the arguments made in our discussion so far, an important process was still omitted. Since we are interested in perturbations whose radial wavelengths are short, the thermal diffusion in the radial direction cannot be neglected. That is, on the right-hand side of energy equation (12), the term of vertically integrated $\text{div}(K \text{grad} T)$ should be added, where $K$ is the thermal conductivity and $T$ the temperature. Since turbulence is considered here as the origin of viscosity, $K/C_p$ ($C_p$ is the specific heat of constant pressure) must be at least of the order of the eddy viscosity, i.e., $K/C_p \sim \alpha_* \rho c_s H \sim \alpha_* \Omega \rho H^2$. Hence, the vertical integration of $K$ is of the order of $\alpha_* \Omega \Sigma H^2 (W/\Sigma T) = \alpha_* \Omega H^2 W/T$. When perturbations with a short wavelength in the radial direction are considered, the perturbed part of the vertical integration of $\text{div}(K \text{grad} T)$ is of the order of $-\alpha_* \Omega (kH)^2 W (T_1/T_0)_0$, where $T_1$ is the temperature perturbation over the unperturbed one $T_0$, and the subscript 0 outside the parentheses means the values at the equator. Hereafter we write the quantity as $-\bar{\alpha}_* \Omega (kH)^2 W (T_1/T_0)_0$, where $\bar{\alpha}_*$ is of the order of $\alpha_*$. If the vertical hydrostatic equilibrium is assumed even in the perturbed state, we obtain from the variation of equation of state

$$\left(\frac{T_1}{T_0}\right)_0 = C_3 w + (C_4 - 1)\sigma. \tag{50}$$

The above consideration shows that in the present case equation (41) can still keep its form, if $\bar{G}_w$ and $\bar{G}_\sigma$ are modified as

$$\bar{G}_w = \frac{Q^+}{W}\left(\frac{\partial \ln \eta}{\partial \ln W}\right)_\Sigma - \frac{Q^-}{W}\left(\frac{\partial \ln Q^-}{\partial \ln W}\right)_\Sigma - C_3 \frac{Q^-_{\text{ad}}}{W} - C_3 \bar{\alpha}_* \Omega (kH)^2, \tag{51a}$$

$$\bar{G}_\sigma = \frac{Q^+}{W}\left(\frac{\partial \ln \eta}{\partial \ln \Sigma}\right)_W - \frac{Q^-}{W}\left(\frac{\partial \ln Q^-}{\partial \ln \Sigma}\right)_W - C_4 \frac{Q^-_{\text{ad}}}{W} - (C_4 - 1)\bar{\alpha}_* \Omega (kH)^2. \tag{51b}$$

The last terms of equations (51a) and (51b) are added to equations (42b) and (42c), respectively. Local perturbations in geometrically thick disks imply $(kH)^2 \gg 1$, and therefore the added terms are the main terms of $\bar{G}_w$ and $\bar{G}_\sigma$.

Again, equations (39) and (40) suggest $O(v) = (kr)^{-1}O(\sigma)$ and $O(w) = (kr)^{-1}O(\sigma)$. Also, $O(n_*) = \bar{\alpha}_*(kH)^2$ implied from the final results and equation (16) leads to $O(u) = (kr)O(\sigma)$. In summary we have

$$O(w) \sim O(v) < O(\sigma) < O(u). \qquad (52)$$

On the right-hand side of energy equation (41), $\bar{G}_\sigma$ is larger than $\bar{G}_u$ by a factor $(kH)^2$. This means that in equation (41) $\bar{G}_u u$ can be neglected in comparison with $\bar{G}_\sigma \sigma$, although $u$ is larger than $\sigma$. A similar argument can be made to neglect the terms of $\bar{G}_v v$ and $\bar{G}_w w$. These considerations lead us to the following approximate set of equations:

$$n_* \sigma - ikr\frac{U}{\Omega r} u = 0, \qquad (16)$$

$$-ikrw - \frac{d\ln(W\Omega_K)}{d\ln r}\sigma = -\alpha_*(kr)^2 u, \qquad (44)$$

$$[n_* + \alpha_*(kH)^2]v + \frac{\kappa^2}{2\Omega^2}\frac{U}{\Omega r}u + i\alpha_* kr\left(\frac{c_s}{\Omega r}\right)^2 \frac{d\ln\Omega}{d\ln r}\left(\frac{\partial\ln\eta}{\partial\ln\Sigma}\right)_W \sigma = 0, \qquad (53)$$

and

$$n_*\Omega(-C_2\sigma) = \bar{G}_\sigma \sigma. \qquad (54)$$

Equation (54) shows that the condition of instability is

$$\bar{G}_\sigma < 0, \qquad (55)$$

and the growth rate is given by $-\bar{G}_\sigma/C_2$, where $\bar{G}_\sigma$ is given by equation (51b), not equation (42c). The remaining equations (16), (44), and (53), are subsidiary relations which determine $u$, $v$, and $w$ for given $\sigma$ and $n_*$. Inequality (55) implies that the condition of instability is practically $C_4 > 1$, since $(kH)^2 > 1$. This condition is equivalent to

$$\beta < \frac{1}{3}. \qquad (56)$$

Thus, optically thick advection-dominated disks (which have $\beta \approx 0$) are unstable, while optically thin ones (which have $\beta \approx 1$) are stable in the limit of short wavelength perturbations.

The cause of this instability can be understood as follows. Let us assume a surface density decrease at a certain radius. If $C_4 > 1$, that leads to a temperature decrease [see equation (50)]. Then, heat flows to the place by thermal diffusion process, leading to an increase of entropy there. Since $S_1 \propto C_1 w - C_2 \sigma$ (where $S_1$ is the entropy change), an entropy increase brings about a decrease of the surface density. This acts to amplify the initial surface density decrease and sets up an instability. The above argument shows that the essential point of the instability is a positive correlation between the changes of temperature and surface density, i.e., $C_4 > 1$. This positive correlation is realized in the radiation pressure dominated disks. In that case, since the hydrostatic balance in the vertical direction implies $p/\rho \propto W/\Sigma \propto H^2$, we have $T \propto p^{1/4} \propto (\Sigma H)^{1/4} \propto [\Sigma(W/\Sigma)^{1/2}]^{1/4} \propto W^{1/8} \Sigma^{1/8}$. So the changes of temperature and surface density occur in the same direction, implying $C_4 > 1$. On the other hand, in the case of no radiation pressure we have $T \propto p/\rho \propto W/\Sigma$, and the variations of $T$ and $\Sigma$ occur with the opposite sign, i.e., $C_4 < 1$.

## 6. Discussion

We have discussed the thermal instability of advection-dominated disks against local perturbations. In the case $\alpha$-viscosity, the disks are always unstable [see equation (36)], while in the case a diffusion-type viscosity they are unstable if a certain condition is satisfied [see equation (49) or equation (55)]. The essence of this instability is related to the fact that such disks have a non-negligible pressure force compared with the centrifugal and gravitational forces. Because of the pressure force, the gas can expand (or contract) in the radial direction when a perturbation is imposed. If the radial wavelength of the perturbation is short, we can expect a large density change by the radial expansion (or contraction). This can occur without much pressure change. That a large density change associated with a small pressure change is the essential ingredient of the instability. The energy sources giving rise to the instability are the unbalance of the heating and the cooling rates caused by a density change and a spatial shear variation associated with the perturbation.

We should emphasize here that we have examined the thermal instability of disks against local perturbations. If global perturbations are considered, situations can be different. There is a possibility that the disk is stable against global perturbations. In the case of flickers in shell burning stars, the stars are unstable only against perturbations localized around the shell burning region (e.g., Schwarzschild & Härm 1965; Fujimoto, Sugimoto 1979) and there is no instability for global perturbations. Non-negligible pressure change is associated with a global perturbation, and this acts to suppress any further temperature increase in the shell burning region. Similar situations might be expected in geometrically thick disks if global perturbations are considered.

Let us discuss the non-linear growth of perturbations in the case of $\alpha$-viscosity. The growth rate of thermal instability is of the order of $(\alpha\Omega)(kr)^{1/2}(H/r)$. The time interval during which a thermal perturbation propagates the radial distance $r$ is $r/U$. Hence, the growth of perturbations during the time interval is $\sim \exp[(kr)^{1/2}(H/r)^{-1}]$. This suggests that the perturbations certainly grow during the propagation through the disk. A natural question is then, what happens at the non-linear stage. Do the thermal perturbations grow to a large amplitude or saturate with a small amplitude? To answer this question, the behavior of flickers in stars is suggestive. The flickers in shell-burning stars grow as long as the approximation of negligible pressure variation is valid. This approximation is violated first at a stage when the perturbation grows to a large amplitude and thus a pressure change in the shell-burning region occurs by a radial expansion of the whole star. After this the perturbation is damped out and the star returns to almost the same state as before. In other words, a sharp spike-like luminosity change occurs in the Kelvin time scale (i.e., thermal time scale). This sharp luminosity rise and fall is called a flicker. Flickers are repeated semi-regularly in a much longer time interval. We speculate that the local thermal instability discussed in this paper will behave like a flicker. That is, a local perturbation grows to a highly nonlinear stage where the pressure change becomes a non-negligible fraction. If so, we can expect spike-like time variations which will occur at various places of disks almost randomly in the thermal time scale. These perturbations propagate in the radial direction. This is interesting in relation to the observational facts that the rapid time variations of X-ray stars can be regarded as ensemble of shots (Negoro, Miyamoto, & Kitamoto 1994).

The above arguments represent a picture of rather violent time variations. The $\alpha$-model, however, may be irrelevant to represent the real situation, since the effects of a spatial variation of angular momentum distribution on the variation of stress tensor is represented only partially in this model. A diffusion-type viscosity will be better to represent the stress tensor due to turbulence. If we take a diffusion-type stress tensor, the growth rate of perturbations is of the order of $\alpha\Omega$ as discussed in section 5.1, which is smaller than that in the case of the $\alpha$-model. In this case, $\alpha\Omega$ times $r/U$ is about $(H/r)^{-2}$, and perturbations may not grow much during their propagation through the disk. If so, the picture that the disks are covered randomly by small amplitude perturbations will be relevant. If the thermal diffusion process is considered, however, the growth becomes again rapid for optically thick advection-dominated disks. In optically thin advection-dominated disks local perturbations are damped.

Finally, we note limitations of our analyses and application of our results. The analyses in this paper have been made by using the vertically integrated equations. The use of such equations will be valid as the first approximation, if a) perturbations have no node in the vertical direction and b) the time scales associated with the perturbations are longer than the time scale of hydrostatic balance. These requirements result in a constraint on the viscosity parameter $\alpha$, as discussed in section 2.1. To judge how much the approximation is really valid, however, a more careful consideration may be necessary, since we treat the case where the vertical scale of disks is much longer than the radial size of perturbations. To obtain a definite conclusion concerning this point, a comparison with two-dimensional analyses will be necessary. If $\alpha$ is not small, two-dimensional analyses will be necessary from the beginning in order to get reliable results. Stability against global perturbations is not examined in this paper. Furthermore, two temperature advection-dominated disks are also outside the scope of this paper.

S. K. thanks Göteborg University and Chalmers University of Technology for their kind hospitality in 1994 May and June when this work was started.